\documentclass[twocolumn,showkeys,preprintnumbers,amsmath,amssymb]{revtex4}
\usepackage{graphicx}
\usepackage{bm}

\usepackage{psfrag}  \usepackage{amsmath,amsthm} \usepackage{bm}

\begin{document}
\def\P{\mathbf{P}} \def\Q{\mathbf{Q}}


\title{Emergence of Zipf's Law in the Evolution of  Communication}
\author{Bernat  Corominas-Murtra$^1$,  Jordi  Fortuny  Andreu$^2$  and
  Ricard V. Sol\'e$^{1,3,4}$}
\affiliation{$^1$ ICREA-Complex Systems  Lab, Universitat Pompeu Fabra
  -Parc de Recerca Biom\`edica de Barcelona (PRBB). Dr Aiguader 88,  08003 Barcelona, Spain\\ $^2$Centre de
  Ling\"u\'istica  Te\`orica  (CLT), Facultat  de  Lletres, Edifici  B,
  Universitat Aut\`onoma  de Barcelona, 08193  Bellaterra (Barcelona),
  Spain\\  $^3$Santa Fe  Institute, 1399  Hyde Park  Road,  New Mexico
  87501, USA \\
  $^4$Institut
  de Biologia Evolutiva. CSIC-UPF.  Passeig Mar\'itim de la Barceloneta,
  37-49, 08003 Barcelona, Spain.}

\begin{abstract}
Zipf's law seems to be ubiquitous in human languages and appears 
to be a universal property of complex communicating systems. Following the 
early proposal made by Zipf concerning the presence of a tension between 
the efforts of speaker and hearer in a communication system, we 
introduce evolution by means of
 a variational approach to the problem based on Kullback's 
Minimum Discrimination of Information  Principle. Therefore, using a formalism fully embedded in the framework of information theory, 
we demonstrate 
that Zipf's  law is the only expected outcome of an evolving, communicative system under a rigorous definition of the communicative tension described by Zipf.
\end{abstract}



\maketitle

\section{Introduction}
Zipf's law is one  of the  most common power laws found in
nature and society \cite{Auerbach:1913, Zipf:1936, Zipf:1949, Bak:1987,
  Gabaix:1999, Newman:2005}. Although it was early observed in the distribution of
money income \cite{Pareto:1896}  and city sizes \cite{Auerbach:1913},
it was popularized by the linguist  George Kingsley Zipf, who observed that it
accounts   for   the  frequency   of   words   within  written   texts
\cite{Zipf:1936,  Zipf:1949}.   Specifically,   if  we  rank  all  the
occurrences of words in a text  from the most common to the least, 
Zipf's law states  that the probability $q(s_m)$ that in a random trial we 
find the $m$-th most common  word ($i=1,...,n$) falls
off as
\begin{equation}
q(s_m) = {1 \over Z} m^{-\gamma},\nonumber
\label{PLAnsatz}
\end{equation}
where
\[
Z=\sum_{j\leq n} j^{-\gamma},
\]
 with $\gamma \approx  1$. The ubiquity of  this scaling
behavior suggested several mechanisms to account for the emergence  of this
distribution,  among  many  others,  see  \cite{Simon:1955,  Bak:1987,
  Li:1992, Harremoes:2001, Ferrer:2003, Corominas-Murtra:2010}.
\begin{figure}[h]
\includegraphics[width=8.5cm]{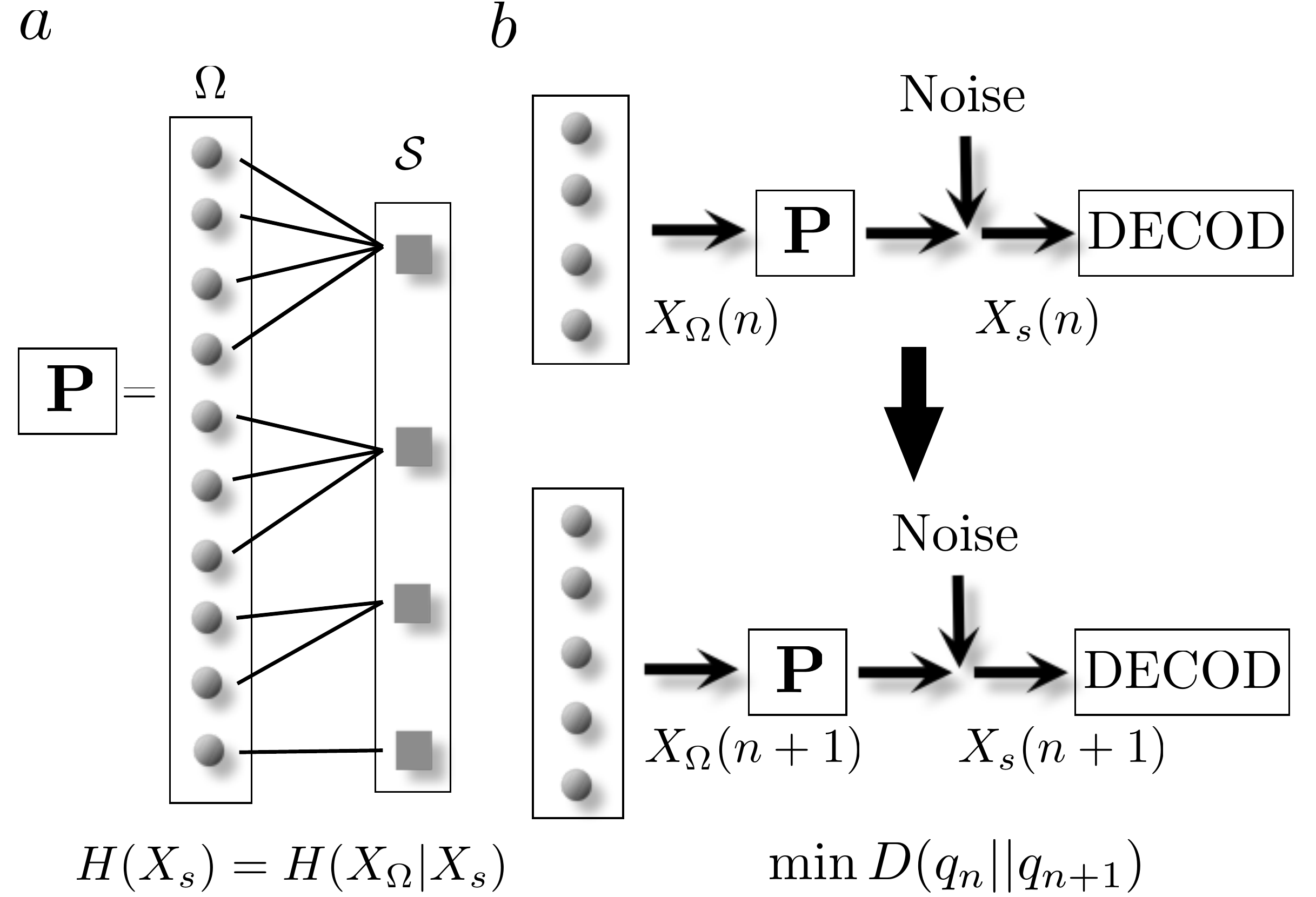}
\caption{A  growing  communication  system.   In  (a)  possible
  meaning-signal associations made by the coder module $\mathbf{P}$ 
  in which eq. (\ref{Symmetryeq}) holds  is
  depicted. In   (b)  we   summarize   the  evolution   rules  of   our
  communicative system. Suppose that symmetry between coder and decoder -i.e., eq. (\ref{Symmetryeq})- 
  holds for the step $n$ (above). At each step
  (below)  a  new element  is  added to  the  set  ${\Omega}$ and  eq. (\ref{Symmetryeq}) holds again for this new configuration. 
  Furthermore, the new configuration is constrained by the $MDIP$, which introduces a path dependency in the evolutionary process. }
\label{zipf1}
\end{figure}

Within the context of human language, G. K. Zipf early conjectured   that 
this scaling law is the outcome of a tension between two {\em forces} acting 
in a communication system \cite{Zipf:1949}. Following Zipf's proposal, speakers and hearers 
need to simultaneously minimize their efforts, under what he called {\em vocabulary balance}, a particular case of the so-called 
{\em  Principle of Least  Effort}. This triggers a tension between the two communicative agents, 
while trying to simultaneously minimize their efforts. The speaker's economy would favour a reduction of the size of the vocabulary to a
single word  whereas  the hearer's economy would lead to an 
increase of the size  of a vocabulary to a  point where there would be
a different {\em word} for each {\em meaning}. The resulting vocabulary would emerge out of 
this unification-diversification conflict \cite{Zipf:1949}. 
Although both numerical and theoretical studies have 
explored this idea \cite{Harremoes:2001, Ferrer:2003, Vogt:2004}, 
no truly analytic proof of unicity has been provided
under realistic, information-theoretic constraints. We  can view  
the proposals made in \cite{Harremoes:2001, Ferrer:2003, Vogt:2004}
  as {\em static}  for they consider  a fixed
size of the code. 

A recent approach -which goes beyond the communicative framework- defined the key complexity properties of a system to display a statistics of events following Zipf's law: An open, unbounded number of accessible states and a linear loss of entropy due to generic internal constraints  \cite{Corominas-Murtra:2010}. The linear loss of entropy grasps the intuitive idea that the studied systems are in an {\em intermediate state} between order and disorder -or that a possible  informative tension is balanced, as we shall see- and the unbounded number of accessible states reflects their open nature. It was shown that, under a very general parametrization, and imposing properties of scale-invariance to the solution, Zipf's law was the only possible outcome.

Now we adapt and enrich the general framework proposed in \cite{Corominas-Murtra:2010} to the communicative context. As we shall see, Zipf's hypothesis can be interpreted in such a way that the system can be studied within the framework proposed in \cite{Corominas-Murtra:2010}. Moreover, the parameters that were arbitrary in the general mathematical framework mentioned above can now be naturally interpreted  in the communicative framework as the key pieces of the mathematical statement of Zipf's hypothesis. 

Beyond the mathematical formalization of the communicative conflict described by Zipf, we need another ingredient, pointed out -in a different context- in \cite{Millart:2008}, namely the active role played by
the evolutionary path  followed by the code. As it occurs with other systems growing out 
of equilibrium, such as scale-free networks \cite{Dorogovtsev:2003}, 
we will consider the evolution of the communicative 
exchange under system's growth. 
Here the evolutionary component is variationally introduced by minimizing the divergence
between code configurations belonging to successive time steps. 
This  minimal change follows the so-called 
{\em Minimum  Discrimination Information Principle} (henceforth $MDIP$), a general variational 
principle considered  analogous to the 
Maximum Entropy Principle \cite{Jaynes:1957}, 
from which statistical mechanics can be properly formalized \cite{Kullback:1959, Plastino:1997}.
The $MDIP$ states
that,  under  changes in  the  constraints  of  the system,  the  most
expected   probability  distribution   is  the   one   minimizing  the
Kullback-Leibler divergence (also referred as to {\em Kullback-Leibler entropy} or {\em relative entropy}) 
from the original one \cite{Kullback:1959}.
Such a  variational  principle constrains  the
changes  of the  internal configurations  of an statistical ensemble when the
external conditions  change in the same way that internal configurations of an statistical ensemble change 
when we introduce moment constraints in a Jaynesian formalism. In our context,
this information theoretic functional assumes the role of a Lagrangian 
whose minimization along the process defines the possible ensemble configurations one can observe 
at a certain point of an evolutionary path.

Using  the $MDIP$ and the framework provided in  \cite{Corominas-Murtra:2010}, we provide a proof of
unicity for the emergence of Zipf's law in evolving codes. We stress that no arbitrary assumptions are made on the nature of solutions.

The remainder of the paper is structured as follows: In section II we rigorously define the communicative tension intuitively defined by Zipf and explicitly characterize the evolutionary process in terms of the mathematical statement of such a tension. In section III we apply the $MDIP$ as the guiding, variational principle which accounts for the possible evolutionary paths of the code. Finally, we demonstrate that the consequences of the application of both the communicative tension and the $MDIP$ account for the emergence of Zipf's law as the unique possible solution of the evolving code. In section IV we discuss the implications of our results.

\section{The evolution of the communicative system}
In this section we mathematically define $1)$ the communicative tension described by Zipf and $2)$ the evolution or growth of a given code subject to such a tension. We furthermore define the range of application of our formalism. As we shall see in section III, the proposal made in this section defines a framework whose key piece to work with is eq. (\ref{UnboundednessXs}).

\subsection{The explicit description of the communicative conflict}
The first task is to properly define the communicative tension between
the coder  and the  decoder  and how  this
tension is solved. Following the standard nomenclature used in studies
of    the    evolution    of    communicating,    autonomous    agents
\cite{Hurford:1989,  Nowak:1999, Komarova:2004},  in our  system there
are two agents: the coder agent, $\P$, encoding information from a set
of external  events, $\Omega$, and  the decoder or  external observer,
which infers the behavior of $\Omega$ through the code provided by the
coder agent $\P$. In this  way, 
\[
\Omega=\{m_1,...,m_n\}
\]
 is the set of
external   events   acting  as   the   input   alphabet,  and   
\[
{\cal S}=\{s_1,...,s_n\}
\]
 is  the set of  signals or output  alphabet. The
coder  module $\P$ -fig. (\ref{zipf1}a)- is  fully  described by a matrix $\mathbb{P}(X_s|X_\Omega)$, 
where  $X_{\Omega}$ is  a random  variable  taking values  on the  set
$\Omega$  following the  probability measure  $p$; being  $p(m_k)$ the
probability  to   have  symbol   $m_k$  as  the   input  in   a  given
computation. Complementarily, $X_s$ is a random variable taking values
on ${\cal  S}$ and following  the probability distribution  $q$ which,
for a given $s_i\in {\cal S}$, reads:
\begin{equation}
q(s_i)=\sum_{k\leq n}p(m_k)\mathbb{P}(s_i|m_k),
\label{nuA}
\end{equation}
i.e.,  the probability  to  obtain  $s_i$  as the  output of  a
codification. We assume that  
\[
(\forall      m_k     \in      \Omega)  \sum_{i\leq    n}\mathbb{P}(s_i|m_k)=1.
\]
For the decoder agent inferring the input set from the output set 
with least effort, the
best scenario is a one-to-one mapping between
${\Omega}$ and ${\cal S}$. In this case, $\P$ generates an unambiguous
code,  and  no supplementary  amount  of  information to  successfully
reconstruct ${X_\Omega}$ is required. However, from the coding device
perspective, this coding has a high cost. In order to characterize this 
conflict, let us properly
formalize the  above intuitive statement:  The decoder agent  wants to
reconstruct  $X_{\Omega}$  through the  intermediation  of the  coding
performed by $\P$.  Therefore, the  amount of {\em bits} needed by the
decoder of $X_s$ to unambiguously reconstruct $X_{\Omega}$ is
\begin{equation}
H(X_{\Omega}, X_s)=-\sum_{i\leq n} \sum_{k\leq n}\mathbb{P}(m_i, s_k)\log \mathbb{P}(m_i, s_k),\nonumber
\end{equation}
which is the {\em joint Shannon entropy} or, simply, {\em joint entropy} of the two random variables $X_{\Omega}, X_s$ \footnote{Throughout the paper, $\log\equiv \log_2$.}. 
From the codification process, the decoder receives $H(X_s)$ bits, and
thus, the remaining uncertainty it must face will be
\[
H(X_{\Omega}, X_s)-H(X_s)=H(X_{\Omega}|X_s),
\]
where
\[
H(X_s)=-\sum_{i\leq n}q(s_i)\log q(s_i),
\]
(i.e, the {\em entropy} of the random variable $X_s$) and
\[
H(X_{\Omega}|X_s)=-\sum_{i\leq n}q(s_i)\sum_{k\leq n}\mathbb{P}(m_k|s_i)\log \mathbb{P}(m_k|s_i),
\]
the {\em conditional entropy} of the random variable $X_{\Omega}$ conditioned to the random variable $X_s$.
The tension between the  coder and the decoder is solved by imposing a 
 symmetric  balance  between its  associated  {\em efforts} -see fig. (\ref{zipf1}a)-, i.e.:  The
 coder sends as many bits as  the additional bits the observer needs to
 perfectly reconstruct $X_{\Omega}$:
\begin{equation}
H(X_s)=H(X_{\Omega}|X_s).
\label{Symmetryeq}
\end{equation}
The above {\em ansatz} is the mathematical formulation of the symmetric balance 
between the efforts of the coder and the decoder. We will refer to this equation as the {\em symmetry condition} and, as pointed out in
\cite{Ferrer:2003}, it mathematically describes 
how  the communicative tension 
is solved  by using a cooperative  strategy between the  coder and the
decoder    agents.     It is worth noting that different equations sharing the same spirit were formerly 
proposed, within the framework of the so-called {\em code-length} game \cite{Harremoes:2001}.   
From eq. (\ref{Symmetryeq}), we can state that:
\[
H(X_{\Omega}, X_s)=2H(X_s).
\]
And knowing the classical inequalities
\begin{eqnarray}
&&H(X_{\Omega},X_s)\geq H(X_{\Omega})\nonumber\\
&&H(X_{\Omega}|X_s)=H(X_{\Omega},X_s)- H(X_s)\leq H(X_{\Omega}),\nonumber
\end{eqnarray}
 we  reach  a general  relation
between the informative richness  of the input variable $X_\Omega$ and
the   informative  richness   of the messages sent  by   the  coder,   constrained  by
eq. (\ref{Symmetryeq}):
\begin{equation}
\frac{1}{2}H(X_{\Omega})\leq H(X_s)\leq H(X_{\Omega}).
\label{GeneralX(s)}
\end{equation}
The  first  relation  becomes  equality  only  in  the  case  of  $\P$
performing a deterministic  codification process.  The second relation
becomes  equality when  the coding  device performs  completely random
associations.  It is clear that eqs.  (\ref{Symmetryeq}) and (\ref{GeneralX(s)}) 
alone cannot explain the  emergence of  Zipf's law  since one
could  tune the  parameters of,  say, an  exponential  distribution to
reach  the desired  relation  between entropies.  Therefore we need to introduce another ingredient
 to obtain Zipf's law as the unique possible  solution of our problem.

\subsection{Evolution}
\label{Evolution}
The  unicity in  the
solution is  provided by  the evolution, which is now explicitly  introduced -see fig (\ref{zipf1}b).  Let us suppose that
our communicative success 
grows over time, thereby increasing the number of input symbols that ${\P}$ can encode.
Formally,  this implies  that the  cardinality of  the  set ${\Omega}$
defined  above increases.  We  introduce this  feature  by defining  a
sequence  of $\Omega$'s  $\Omega(1), ...,\Omega(k),...$  satisfying an
inclusive ordering, i.e.,
\[
\Omega(1)\subset \Omega(2)\subset...\subset\Omega(k),...,
\]
which  is introduced, without  any loss  of generality,  assuming that
\begin{eqnarray}
\Omega(1)&=&\{m_1\},\nonumber\\ 
\Omega(2)&=&\{m_1,m_2\}, \nonumber\\
...&&...\nonumber\\
\Omega(n-1)&=&\{m_1,...,m_{n-1}\},\nonumber\\
\Omega(n)&=&\{m_1,...,m_{n-1},m_n\}.\nonumber
\end{eqnarray}
At time  step $n$, ${\P}$ will be  able to process the  $n$ symbols of
$\Omega(n)$.
The  elements
$m_1,...,m_i,...$  are   members  of  some   infinite,  countable  set
$\tilde{\Omega}$,       i.e.,       $(\forall      i)(\Omega(i)\subset
\tilde{\Omega})$.   $\tilde{\Omega}$   can  be  understood,   using  a
thermodynamical  metaphor,  as   a  {\em  reservoir  of  information}.
Following this characterization, we  say that for every set $\Omega(i)$
there  is   a  random  variable  $X_{\Omega}(i)$,   taking  values  in
$\Omega(i)$  following  the  ordered probability  distribution  $p_i$.
Furthermore,  we  assume  that exists a unique $ \mu\in  (0,1)$  such  that
$(\forall \epsilon >0)(\exists N):(\forall n>N)$,
\begin{equation}
\left| \frac{H(X_{\Omega}(n))}{\log n}-\mu\right|<\epsilon.
\label{Unboundedness}
\end{equation}
This means that  the entropy of  the input set  is unbounded when  its size
increases, which implies that the potential input set $\tilde{\Omega}$
acts as an {\em infinite} reservoir of information.
  The behavior of the output set at the stage $n$ is  described by a random
variable $X_s(n)$,  which follows the  ordered probability distribution
$q_n$,  as  defined  in  eq.   (\ref{nuA}), taking  values  on  ${\cal
  S}(n)=\{s_1,...,s_n\}$. We observe  that ${\cal S}(n)\subseteq {\cal
  S}(n+1)$, defining a sequence ${\cal S}(1),...,{\cal S}(k),...$ also
ordered  by inclusion.  At every  time step,  the consequences  of the
symmetry   condition   -see   eq.  (\ref{Symmetryeq})-   depicted   in
eq. (\ref{GeneralX(s)}) are satisfied, which implies that the sequence
\[
{\cal H}=H(X_s(1)),    H(X_s(2)),...,H(X_s(k)),...
\]
  also   satisfies    the
convergence ansatz  made over the sequence of  normalized entropies of
the input -see eq.  (\ref{Unboundedness}).  The only difference is the
value  of the  limit, $\nu$. The value of $\nu$ can be bounded by using eqs (\ref{GeneralX(s)}) and (\ref{Unboundedness}), thereby obtaining:
\begin{equation}
\frac{1}{2}\mu\leq \nu\leq\mu
\label{1/2mu}
\end{equation}
Therefore, in  this case,  by  virtue of
eqs. (\ref{GeneralX(s)}), (\ref{Unboundedness}) and (\ref{1/2mu}), the convergence
condition  for the  normalized  entropies of  the  sequence of  random
variables    $X_s(1),...,     X_s(n),...$    reads: exists a unique    $\nu\in
(\frac{1}{2}\mu,  \mu)$  such   that  $(\forall  \epsilon  >0)(\exists
N):(\forall n>N)$:
\begin{equation}
\left| \frac{H(X_{s}(n))}{\log n}-\nu\right|<\epsilon.
\label{UnboundednessXs}
\end{equation}
The  above  equation depicts  two  crucial  facts  in the  forthcoming
derivations: If the potential informative richness of the input set is
unbounded, so is the informative richness of the output set, under the
constraints  imposed  by  the symmetry condition -see eq.  (\ref{Symmetryeq}).
 \section{The emergence of Zipf's Law under the $MDIP$}
 
 The $MDIP$ is presented in this section as the variational principle guiding the evolution of the code. As we shall see at the end of this section, the consequences of its application result in a proof of unicity for the emergence of Zipf's law in evolving codes.
 
 \subsection{The MDIP and its consequences for the evolution of codes}
 
The question  is thus how  the probability distribution  $q_n$ evolves
along  the  growth  process.   Under the $MDIP$ we face a variational problem which 
is stated as follows: During  the  growth
process, the most likely code at  step $n+1$ is the one minimizing the
{\em  distance} with  respect  to the  code  at step  $n$, consistently with 
the $MDIP$. Furthermore, the evolution of the code must satisfy, along all the 
evolutionary steps, the symmetry condition depicted by eq. (\ref{Symmetryeq}). The  crucial 
contribution of the $MDIP$ is that it naturally introduces the footprints of the path dependence imposed by evolution.
Following  the thermodynamical  metaphor,  this variational  principle
acts, in  our context,  as a principle  on energy  minimization acting
over  the transitions of  successive codes.   
Putting it formally, let 
\begin{equation}
D(q_{n}||q_{n+1})                                               
\equiv \sum_{i\leq n+1}q_{n}(s_i)\log\frac{q_{n}(s_i)}{q_{n+1}(s_i)}\nonumber
\end{equation}
be the {\em Kullback-Leibler} Divergence of the distribution $q_{n+1}$
with  respect the  distribution $q_n$  \cite{Thomas:2001}.  Therefore,
the  $MDIP$ is achieved by  minimizing the  following functional \cite{Kullback:1959}:
\begin{equation}
{\cal L}(q_{n+1},  \lambda)=D(q_{n}||q_{n+1})+\lambda\left(  \sum_{i\leq n+1}
q_{n+1}(s_i)-1\right).\nonumber
\label{KullbackFunct}
\end{equation}
We observe that this functional has a role equivalent to the one attributed to the  Lagrangian function in a given continuous, differentiable system; therefore, the {\em trajectories} minimizing it will govern the evolution of the system. 
Furthermore, the  symmetry condition 
on  coding/decoding -eq. (\ref{Symmetryeq})- imposes that
the    solutions    must    lie    in   the    region    defined    by
eq.  (\ref{UnboundednessXs}).   The  minimum  of  ${\cal L}$  is  found  when
$q_{n+1}$ satisfies:
\begin{equation}
	q_{n+1}(s_i)=\left\{
	\begin{array}{ll}
		\frac{1}{\lambda}    q_n(s_i)\;\;{\rm    iff}\;i\leq    n\\    
		1-\frac{1}{\lambda}\;\;{\rm  iff}\;i=n+1,
	\end{array}
	\right.
\label{elambdaDistrib}
\end{equation}
being $\lambda$  the Lagrange multiplier, which is  a positive, unique
constant for  all elements of the  probability distribution $q_{n+1}$.
We observe that, for  $\lambda=1$, $D(q_{n}||q_{n+1})=0$, but, in this
case,  $H(X_s(n))=H(X_s(n+1))$,  in  contradiction to  the  assumption
provided   by   eq.   (\ref{UnboundednessXs}),  according   to   which
informative richness grows during the evolutionary process.
\begin{figure}
\includegraphics[width=7.5cm]{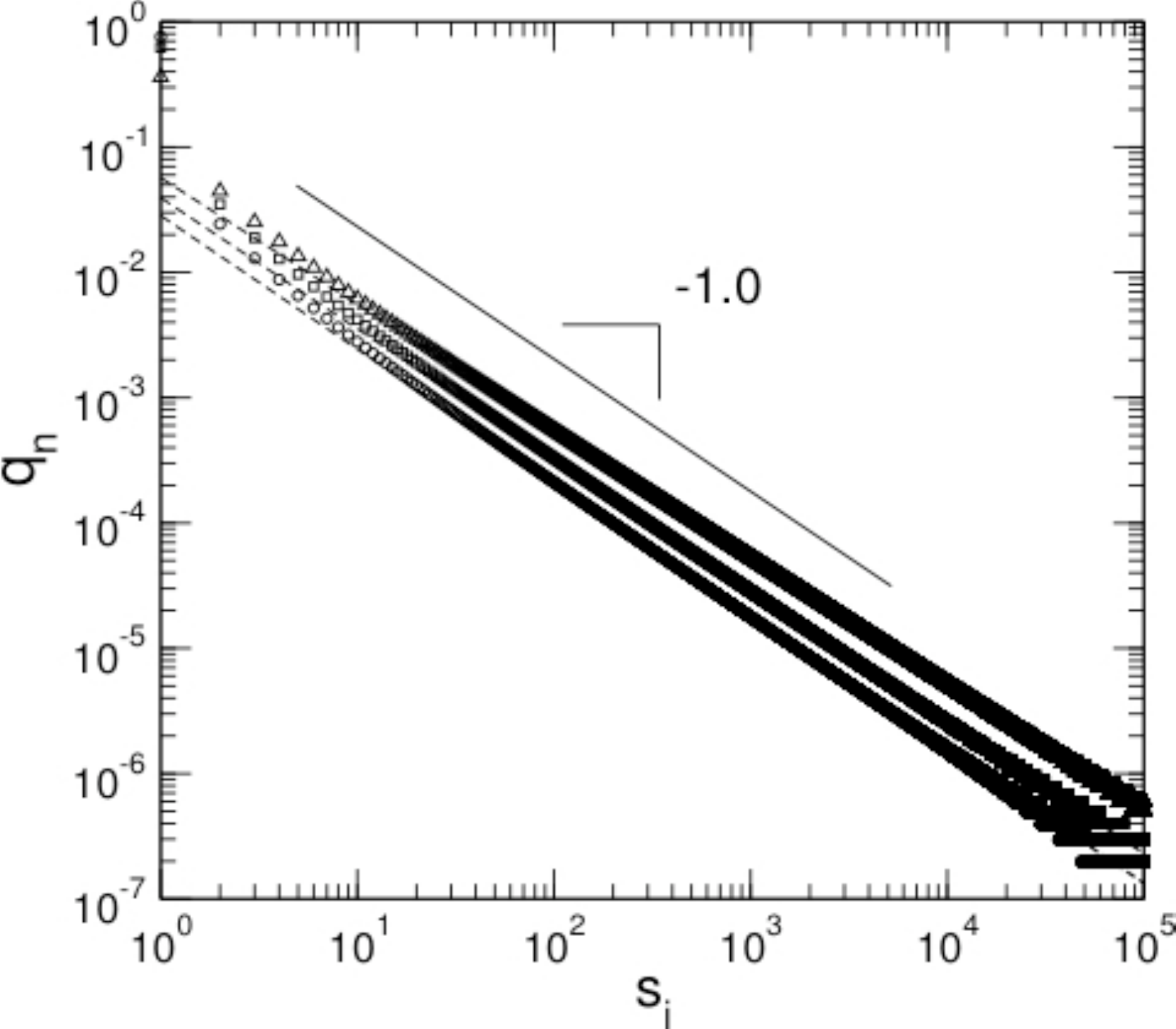}
\caption{Numerical  simulation   of  the  final   distribution  $q_n$
  ($n=10^4$) obtained  by constraining the growth  process with i)the
  consequences    of   the    symmetry    of   coding/decoding    -see
  eq. (\ref{Symmetryeq})- provided  by eq. (\ref{UnboundednessXs}) and
  ii) the  application  of the  $MDIP$  at  every  step of  the  growth  process. 
  Different convergence values  are studied: a) $\nu=0.2$, b)
  $\nu=0.3$ and c) $\nu=0.5$. The dashed lines are the best fit interpolations, 
which give estimated exponents $\gamma=1.06, 1.04$ and $1.01$, respectively 
(all with correlation coefficients $r<-0.99$). }
\label{zipf2}
\end{figure}

Now we  want to find the  asymptotic behavior of  $q_n, \;n\to \infty$
under  the  above  justified  conditions  (\ref{UnboundednessXs})  and
(\ref{elambdaDistrib}).   The  key   feature  we derive  from  the  path
dependency  in  the  evolution  imposed  by the  $MDIP$  is  that  the
following quotient
\begin{equation}
(\forall k+j\leq n)\;\;f(k,k+j)=\frac{q_n(s_{k+j})}{q_n(s_{k})}
\label{f(m,m)}
\end{equation}
does not depend on $n$.  Therefore, along the evolutionary process, as
soon  as 
\[
q_n(s_k),q_n(s_{k+j})>0,
\]
$f(k,k+j)$ remains  invariant. 

\subsection{The Emergence of Zipf's Law}
The asymptotic behavior
of  quotient $f$  and,  thus, the  tail  of  $q_n$, is  strongly
constrained    by     the    entropy    restriction     provided    by
eq.   (\ref{UnboundednessXs})   \cite{Corominas-Murtra:2010}.   
As shall see, the key of the forthcoming derivations will be the convergence properties of the normalized entropies of a given random variable $X$ having $n$ possible states whose (ordered) probabilities follow a power-law distribution function, namely $g(s_i)\propto i^{-\gamma}$. The explicit form of these entropies is:
\begin{equation}
\frac{H(X)}{\log n}=\frac{1}{\log n}\left(\frac{\gamma}{Z_{\gamma}}\sum_{i\leq n} \frac{\log i}{i^{\gamma}}+\log Z_{\gamma}\right).
\label{H(Y)}
\end{equation} 
Consistently, $Z_{\gamma}$ is the normalization constant.

The first observation is that it can be shown that the convergence properties of the Riemann   $\zeta$-function   on
$\mathbb{R}^+$  \cite{Abramowitz:1965}
\[
\zeta(\gamma)=\sum_{i=1}^{\infty}\frac{1}{i^{\gamma}},
\]
strongly constrain the convergence properties of a given probability distribution \cite{Corominas-Murtra:2010}.
Indeed,  we  find that,  if  $(\forall
\delta >0,\;n>m)(\exists N)$ such that:
\begin{equation}
(\forall
  m>N)\;\;\;f(m,m+1)<\left(\frac{m}{m+1}\right)^{1+\delta},\nonumber
\end{equation}
then
$(\exists   C<\infty  \in   \mathbb{R}^+)\;{\rm  such\;that}\;(\forall n)(H(X_s(n))<C)$,
which  contradicts the  assumptions of  the problem,  depicted  by eq.
(\ref{UnboundednessXs}). 
Indeed, primarily, one can observe that the above statement implies that
$q_n$ is {\em dominated} by a power-law having exponent $1+\delta$, i.e. that $q_n$ decays faster than $q'_n$, defined as:
\[
q'_n(s_i)=\frac{i^{-(1+\delta)}}{Z_{1+\delta}},
\]
where $Z_{1+\delta}$ is the normalization constant. Now, we write the explicit form of the entropy of $X'_s(n)\sim q'_n$ -to be written as $H(X'_s(n))$- 
when $n\to\infty$ by multiplying the expression derived in eq. (\ref{H(Y)}) by $\log n$:
\[
\lim_{n\to\infty}H(X'_s(n))=\frac{1+\delta}{\zeta(1+\delta)}\sum_{i=1}^{\infty}\frac{\log i}{i^{1+\delta}}+\log(\zeta(1+\delta)).
\]
We observe that all the elements of the above equation are finite constants, since
\[
\sum_{i=1}^{\infty}\frac{\log i}{i^{1+\delta}}<\infty.
\]
Thus, having $q'_n$ as defined above,
\[
\lim_{n\to\infty}H(X'_s(n))<\infty.
\]
Therefore,  during
the growth process, due to the constraint imposed by eq. (\ref{UnboundednessXs}),
\begin{equation}
f(m,m+1)>\left(\frac{m}{m+1}\right)^{(1+\delta)},
\label{>1+d}
\end{equation}
with  $\delta$  arbitrarily  small,  provided that  $n$  can  increase
unboundedly.  Otherwise, its normalized entropy -see eq. (\ref{H(Y)})- will have, as an asymptotic value
\[
\frac{H(X_s(n))}{\log n}\to 0,
\]
in contradiction to the assumption that $\nu>0$ as depicted in eq. (\ref{UnboundednessXs}).

Furthermore, we observe that, if $(\forall \delta >0,\;n>m)(\exists
N)$ such that
\begin{equation}
(\forall
  m>N)\;\;f(m,m+1)>\left(\frac{m}{m+1}\right)^{(1-\delta)},
  \label{Dominated}
\end{equation}
then
\begin{eqnarray}
\lim_{n\to \infty}\frac{H(X_s(n))}{\log n}=1,\nonumber
\end{eqnarray}
again in  contradiction to eq. (\ref{UnboundednessXs}),  except in the
extreme, pathological case where $\nu=1$,
when the coding process is completely noisy. To see how we reach this latter point we observe that statement (\ref{Dominated})  implies 
that $q_n$ is {\em not dominated} by a power-law probability distribution 
$q'_n$ having exponent $1-\delta$, namely:
\[
q'_n(s_i)=\frac{i^{-(1-\delta)}}{Z_{1-\delta}},
\]
where $Z_{1-\delta}$ is the normalization constant. Putting explicitly the expression of the normalized entropy -see eq. (\ref{H(Y)})- for the random variable $X'_s(n)$, one obtains:
\begin{eqnarray}
\lim_{n\to\infty}\frac{H(X_s'(n))}{\log n}&=&\lim_{n\to\infty}\left(\frac{\delta(1-\delta)}{n^{\delta}\log n}\sum_{k\leq n}\frac{\log k}{k^{1-\delta}}+\delta\right)\nonumber\\
&=&\lim_{n\to\infty}\frac{1-\delta}{\log n}\left(\log n-\frac{1}{\delta}\right)+\delta\nonumber\\
&=&1,\nonumber
\end{eqnarray}
which is the desired result.
 Accordingly, since from eq. (\ref{UnboundednessXs}) $\nu$ is generally different from $1$,
\begin{equation}
f(m,m+1)<\left(\frac{m}{m+1}\right)^{(1-\delta)}.
\label{<1-d}
\end{equation}

Thus, combining eq. (\ref{>1+d})  and (\ref{<1-d}), we have shown that
the  asymptotic  solution  is   bounded  by  the  following  chain  of
inequalities:
\begin{equation}
\left(\frac{m}{m+1}\right)^{(1+\delta)}<f(m,m+1)<\left(\frac{m}{m+1}\right)^{(1-\delta)}.\nonumber
\end{equation}
The crucial step is that it can be shown that, if $n\to \infty$, we can set
\[
\delta\to 0.
\]
(The mathematical technicalities of this result can be found in \cite{Corominas-Murtra:2010}.)
This implies, in turn, that, for $n\gg 1$:
\[
f(m,m+1)\approx \frac{m}{m+1}
\]
and, from  the definition  of $f$ provided  in eq.  (\ref{f(m,m)}), we
conclude that
\begin{equation}
q_{n}(s_m)\propto \frac{1}{m},\nonumber
\end{equation}
thereby leading  us to  Zipf's  law  as the  unique  asymptotic solution.   
 
In fig. (\ref{zipf2})  we numerically explored  the behavior of  the rank
probability distribution of signals  belonging to a growing code under
the  assumption  of  symmetry  in  coding/decoding  provided  by  eqs.
(\ref{Symmetryeq}) and  (\ref{UnboundednessXs}), and the  $MDIP$ whose
consequences  in   the  evolution  of   $q_n$  are  depicted   in  eq.
(\ref{elambdaDistrib}).    The  outcome   perfectly   fits  with   the
mathematical  derivations, showing  very well-defined  power-laws with
exponents close  to $1$ although the convergence  values $\nu$ diverge
from  $0.2$  to  $0.5$.   This  numerical validation  shows  that  the
predicted asymptotic effects -i.e., the convergence of $q_n$ to Zipf's
law- are perfectly appreciated even in finite simulations where $10^5$
signals are at work.

We end this section with a remark on the boundary conditions needed for the emergence of Zipf's law. In the section \ref{Evolution},
we imposed that the potential information richness of the source must be unbounded. Such a condition is 
mathematically stated by (\ref{Unboundedness}). We observe that, more than an assumption, equation (\ref{Unboundedness}) 
is  a boundary condition under which a growing code can (assymptotically) exhibit Zipf's law\footnote{We notice that eq. (\ref{Unboundedness}) depicts a linear relation between $H(X_{\Omega}(n))$ and $\log n$; i.e.: $H(X_{\Omega}(n))\sim \mu\log n$. There are strong reasons to believe that one could generalize this statement by saying that the only condition needed is that, in spite that 
\[
\lim_{n\to\infty}\frac{H(X_{\Omega}(n))}{\log n}=0,
\]
if $H(X_{\Omega}(n))$ is a monotonous, growing and unbounded function on $n$, then Zipf's law would emerge using similar arguments to the ones used in this paper. The lack of a rigorous demonstration for this latter point forces us to restrict our arguments to the region of application of eq. (\ref{Unboundedness}).}. In this way, since $H(X_s(n))$  has a linear relation with $H(X_{\Omega}(n))$, the divergence of the latter implies the divergence of the former. And it is a required condition, since the entropy of a system exhibiting a power law with an exponent equal to $1$ diverges with $n$. Otherwise, exponents are higher, or other probability distributions can emerge.


\section{Discussion}
The results  provided in our study define a general rationale for the emergence  of Zipf's
law  in the abundance  of signals of evolving communication systems.   The variational 
approach taken here as a formal picture of the least effort hypothesis has two
ingredients.  First, starting  from  Zipf's conjecture,  we reach  a
static symmetry equation  to solve the communicative tension between coder and decoder. 
This is consistent with previous work, but reveals itself
insufficient  to derive Zipf's  law as  the unique solution, for it is easy to check that 
{\em  static}   equations  of  the  kind   of  eq.
(\ref{Symmetryeq})  and (\ref{GeneralX(s)}) have infinite arbitrary solutions,
even in the  asymptotic regime, due to the possible parametrizations of the solutions.
Secondly  -and crucially-  we consider  that the  code  evolves through
time, and  that, consistently, there is  a path dependence  in its evolution,  mathematically  
stated  by imposing a variational principle, the $MDIP$, 
between successive states of  the code. 
It is only  by imposing evolution (and 
thus, path dependence) that we reach Zipf's law as the only asymptotic
solution. Therefore, the  origin of the power-law with exponent
$\gamma=-1$ derives  from three complementary,  very general conditions:
\begin{itemize}
\item
 The  unbounded informative  potential of the  code,  
 \item
 the
loss   of   information   resulting   from   the   symmetry   condition, depicted in eq. (\ref{Symmetryeq}), and
\item
evolution, and its associated path dependency, variationally imposed by the application of the $MDIP$ over successive states of the evolution of the system.
\end{itemize}

There is another, very interesting point,
intimately tied to a code exhibiting Zipf's law and, more specially, the consequences of the {\em symmetry condition}, the mathematical ansatz which abstractly encodes the Zipf's hypothesis of vocabulary balance: The presence of an inevitable ambiguity in the code. It is a common observation that natural languages are ambiguous, namely, that linguistic utterances 
or parts of linguistic utterances can be assigned more than one interpretation. If the principle of least effort is at work, and thus there is a cooperative strategy between the coder and the decoder, then the presence of a certain amount of ambiguity is expected, provided that the speaker tends to assign more than one meaning to certain signals. Therefore, ambiguity is a  byproduct of efficient communication rather than a fingerprint of poor communicative design.

The presented framework is general, and rigorously demonstrates that Zipf's law is a natural outcome of a broad class of communication systems evolving under coding/decoding tensions. In other words, Zipf's law emerges in a system where the coder and decoder {\em coevolve} under a general problem of energy minimization. The range of application to real-world phenomena, however, must be contrasted with the validity of data, for it has been pointed out that many supposed power-law behaviors show deviations when the statistical analysis is performed accurately \cite{Kanter:1995, Avnir:1998}. It should be noted, however, that a deviation of the predicted behavior  need not be necessarily attributed to a failure of the framework. One should take into account that other constraints, such as general memory limitations, can play a role in shaping the final distribution.


\begin{acknowledgments}
We  thank  the   members  of  the  Complex  Systems   Lab  for  useful
discussions and an anonymous reviewer for his/her constructive comments.   
This work has  been supported  by NWO  research project
Dependency  in Universal  Grammar, the  Spanish MCIN  {\em Theoretical
  Linguistics}  2009SGR1079 (JF), the  James S.   McDonnell Foundation
(BCM) and by Santa Fe Institute (RS).
\end{acknowledgments}


\end{document}